\documentstyle[prl,tighten,aps,epsf,graphics,multicol,amstex]{revtex}

\newcommand{\bra}[1]{\langle{#1}|}

\newcommand{\ket}[1]{|{#1}\rangle}
\newcommand{\ip}[1]{\langle{#1}\rangle}
\newcommand{\non}{\nonumber}
\newcommand{\id}{\openone}

\newcommand{\eqr}[1]{Eq.~(\ref{#1})}
\newcommand{\be}{\begin{equation}}
\newcommand{\ee}{\end{equation}}
\newcommand{\bea}{\begin{eqnarray}}
\newcommand{\eea}{\end{eqnarray}}
\newcommand{\beq}{\begin{equation}}
\newcommand{\eeq}{\end{equation}}
\newcommand{\beqa}{\begin{eqnarray}}
\newcommand{\eeqan}{\end{eqnarray*}}
\newcommand{\beqan}{\begin{eqnarray*}}
\newcommand{\eeqa}{\end{eqnarray}}

\begin{document}
\draft
\title{Remote control of restricted sets of operations:\\
Teleportation of Angles}
\author{S.F. Huelga$^1$, M.B. Plenio$^2$ and J.A. Vaccaro$^1$}
\address{$^1$Department of Physical Sciences, University of Hertfordshire, Hatfield AL10 9AB,
UK\\ $^2$QOLS, Blackett Laboratory, Imperial College of Science, Technology and
Medicine, London, SW7 2BW, UK}
\date{\today}
\maketitle
\begin{abstract}
We study the remote implementation of a unitary transformation
on a qubit. We show the existence of non-trivial protocols
(i.e., using less resources than bidirectional state
teleportation) which allow the perfect remote implementation of
certain continuous sets of quantum operations. We prove that,
up to a local change of basis, only two subsets exist that can
be implemented remotely with a non-trivial protocol: Arbitrary
rotations around a fixed direction $\vec{n}$ and
rotations by a fixed angle around an arbitrary direction lying in
a plane orthogonal to $\vec{n}$. The overall classical information
and distributed entanglement cost required for the remote
implementation depends on whether it is a priori known to which of
the two teleportable subsets the transformation belongs to. If it
is so, the optimal protocol consumes one e-bit of entanglement
and one c-bit in each direction. If the subset is not known, two
e-bits of entanglement need to be consumed while the classical
channel becomes asymmetric, two c-bits are conveyed from Alice to
Bob but only one from Bob to Alice.
\end{abstract}
\section{Introduction}
Using entanglement as a resource is a common feature of many tasks
in quantum information processing \cite{review}. A canonical
example of entanglement-assisted processes is provided by quantum
state teleportation \cite{tele}, where an arbitrary qubit state
can be transferred with perfect fidelity among distant parties
with the sole use of two classical bits (c-bits) and the
consumption of a distributed maximally entangled state, i.e., one
e-bit of shared entanglement. Recently we have a addressed a
related problem where the aim is to {\em teleport} across distant
parties not a quantum state but a {\em quantum operation}
\cite{teleU}. By this we mean the following. Alice and Bob are set
in remote locations and one of the parties, say Alice, is given a
black box with the ability of performing a very large set of
unitary transformations $U$ on a qubit. The requirement of the set
of allowed transformations being very large is imposed with the
aim of excluding, by construction, the possibility of teleporting
the full black box to Bob, which would exhaust entanglement
resources very quickly. We will say that the operation
$U$ has been teleported to Bob, or equivalently, it has been
remotely implemented if, for any qubit state Bob may hold, a
protocol involving only local quantum operations and exchange of
classical communication (LQCC) can yield a final global state
where Bob holds the state transformed by the operation $U$
disentangled from any other system (See below for a quantitative
formulation). Our previous results show that if we want the
transformation $U$ to be an arbitrary element of the group $SU(2)$, no LQCC
protocol can exist consuming less overall resources than
teleporting Bob's state to Alice followed by Alice teleporting the
state transformed by $U$ back to Bob. In other words, the remote
implementation of an arbitrary unitary operation on a qubit cannot
be accomplished by means of any local protocol which uses less
resources than bidirectional quantum state teleportation (BQST).
This amounts to two e-bits of entanglement and two classical bit
in each direction. The ultimate responsible for this result is
linearity. Therefore, the impossibility of implementing remotely
an arbitrary $U$ without resorting to state transfer belongs to
the family of no-go results imposed by the linear structure of
quantum mechanics
and exemplified, for instance, by the non-cloning theorem \cite{cloning}.\\
What happens if the requirement of being able to implement {\em
any} $U$ is relaxed? Can we find families of operators that can be
implemented consuming less overall resources than BQST?. We should
stress that we are interested here in exploiting entanglement,
therefore any strategy which attempts the local reconstruction
of $U$ \cite{bcn} is excluded from our valid protocols. In
addition, we want to keep to a minimum the available a priori
information about $U$. Note, in particular, that if both the form
of $U$ and Bob's initial state are completely known, the posed
problem reduces to remote state preparation \cite{remote}.
Finally, we want the procedure to work with perfect efficiency. Imperfect storage of
quantum operations have been
recently discussed by Vidal et al.
\cite{guifre}. We will show that there are indeed two restricted
classes of operations that can be implemented remotely using less
overall resources than BQST and only two (up to a local change of
basis). These are arbitrary rotations around a fixed direction
$\vec{n}$ and rotations by a fixed angle around an arbitrary
direction lying in a plane orthogonal to $\vec{n}$.

We have organized the paper in seven further sections. Section II revises
the necessary resources for achieving the remote implementation of
an arbitrary $U$. In section III a LQCC protocol exhausting these
resources and achieving the maximum probability of success allowed
for arbitrary $U$ is constructed. Remarkably, two possible sets of
transformations could be implemented accurately with this
procedure, as discussed in section IV. A geometrical picture of why it is possible to engineer a
final correction step in these cases is presented in section V.
The uniqueness of the subsets is proven in
section VI, the technical bulk of this paper.
Section VII deals with the resources trade-off when some a priori
information about the functional form of the transformation $U$ is
provided. Final section VIII ends summarizes the results and end up with proposing an
experimental scenario where the teleportation of {\em angles} could be demonstrated.

\section{Remote implementation of an arbitrary $U$: Necessary resources}
Assuming the black box to be a classical system, we are seeking a protocol with
the following structure \cite{teleU}
\begin{equation}
  G_2 \, U \, G_1 (\ket{\chi}_{aAB} \otimes\ket{\psi}_{b})
   =  \ket{\Phi(\chi)}_{aAB} \otimes U \ket{\psi}_{b}
  \label{doble},
\end{equation}
where certain fixed operations $G_1$ and $G_2$ are performed,
respectively, prior to and following the action of the arbitrary
$U$ on a qubit $a$ on Alice's side. The fact that the operation
$G_1$ has to be non-trivial follows from the results of Nielsen
and Chuang when analyzing universal programmable gates
\cite{nichu}. We assume that Alice and Bob share initially some
entanglement, represented by the joint state $\ket{\chi}_{\alpha
AB}$. The purpose of the protocol is to end up with Bob holding a
qubit in the transformed state $U \ket{\psi}_{b}$, for any initial
state $\ket{\psi}_{b}$ and with perfect efficiency. Note that the
final distributed state involving the remaining subsystems $aAB$
is independent of both $U$ and $\ket{\psi}_{b}$ \cite{teleU}. As
in \cite{teleU}, it will convenient to use a nonlocal unitary
representation of the transformation, with $G_1$ and $G_2$ being
unitary operators acting on possibly all subsystems. For instance,
a possible solution, while in principle not necessarily optimal,
corresponds to each $G_i$ being a state teleportation
process.
\noindent In the following we will establish lower bounds on the amount of
classical communication and the amount of entanglement required for the
teleportation of an arbitrary unitary transformation. Our argument employs the
principle that entanglement cannot be increased under LQCC to show that 2
e-bits are necessary and it uses the impossibility of superluminal
communication to demonstrate that 2 classical bits have to be sent from Alice
to Bob and at least one bit has to be transferred from Bob to Alice.\\
Assume that we could teleport any arbitrary operation $U$ from
Alice to Bob. Therefore, a universal protocol involving operations
$G_1$ and $G_2$ would yield the outcome $\ket{\Phi(\chi)}_{aAB}
\otimes U \ket{\psi}_{b}$, independently of the actual form of
$U$. It is easy to show that then it would also be possible to
implement remotely an arbitrary {\em controlled}-U gate. By this
we mean that the remote implementation of $U$ is performed
conditional on the state of certain control qubit $c$, so that the
action of the black box is to apply the identity if the control
qubit is state $\ket{0}_c$ and to apply $U$ when the control bit
is state $\ket{1}_c$. That is, Eq.(\ref{doble}) is replaced by
\begin{equation}
  G_2 \, U_c \, G_1 (\ket{\chi}_{aAB} \otimes\ket{\psi}_{b})
   =  \ket{\Phi(\chi)}_{aAB} \otimes (c_0 \ket{0}_c \otimes \id \ket{\psi}_{b}
   + c_1 \ket{1}_c \otimes U \ket{\psi}_{b})
  \label{dobleU},
\end{equation}
where
\begin{equation}
U_{C} = \ket{0}_{cc}\bra{0} \otimes \id + \ket{1}_{cc}\bra{1} \otimes U
\end{equation}
and $\ket{c}=c_0 \ket{0}_c + c_1 \ket{1}_c$ is an arbitrary state of the
control qubit, which, without loss of generality, can be assumed to be part of
the black box and therefore unaffected by the action of the operations $G_i$,
$(i=1,2)$. Let us decompose the global state after the application of $G_1$ as
follows
\begin{equation}
\ket{c} \otimes G_1 (\ket{\chi}_{aAB}\otimes \ket{\psi}_{b})= (c_0 \ket{0}_c +
c_1 \ket{1}_c) \otimes (\ket{0}_{a} \ket{\xi}_0 + \ket{1}_{a} \ket{\xi}_1)
\end{equation}
where the, possibly distributed, states $\ket{\psi}_i$ are neither necessarily
orthogonal not normalized. The action of $U_c$ brings this state onto
\begin{equation}
U_c(c_0 \ket{0}_c + b \ket{1}_c) \otimes (\ket{0}_{a} \ket{\xi}_0 + \ket{1}_{a}
\ket{\xi}_1)= c_0 \ket{0}_c (\id \ket{0}_{a} \ket{\xi}_0 + \id \ket{1}_{a}
\ket{\xi}_1 ) + c_1 \ket{1}_c (U \ket{0}_{a} \ket{\xi}_0 + U \ket{1}_{a}
\ket{\xi}_1)
\end{equation}
Now, the subsequent action of the operation $G_2$ gives the transformation law
Eq.(\ref{dobleU}) provided that Eq.(\ref{doble}) holds for every qubit
transformation $U$.
\\ A simple controlled-$U$ operation is not yet sufficient for our argument but
we have to introduce a slightly more involved gate. Assume now that we have two
control qubits, c and c', on Alice's side and consider again Bob's qubit as the
target. We will apply a particular operation which we call a controlled Pauli
gate (CP-gate). This gate applies one of the four Pauli-operators on the target
qubit depending on the state of the two control qubits and can be written as
\begin{eqnarray}
U_{CP} &=& \ket{00}\bra{00} \otimes \id + \ket{01}\bra{01} \otimes \sigma_x
\nonumber \\ &+&
 \ket{10}\bra{10} \otimes \sigma_y +
\ket{11}\bra{11} \otimes \sigma_z ,
\end{eqnarray}
where we have omitted the subscripts $cc'$ to make the notation
lighter. Given that we are assuming that Alice can teleport any
unitary operation to Bob, we can therefore implement a CP-gate
between Alice and Bob with Alice acting as the control. We will
demonstrate that the CP-gate can be used to establish, starting
from a product state between Alice and Bob, a state that contains
two shared e-bits. To this end, assume that Bob holds two
particles in the maximally entangled state
$\ket{\phi^+}_B=\ket{00}_B+\ket{11}_B$ and that Alice holds her
two control particles in state
$\ket{00}+\ket{01}+\ket{10}+\ket{11}$. The result of the CP
operation is
\begin{eqnarray}
U_{CP}(\ket{00}+\ket{01}+\ket{10}+\ket{11})_{cc'} \otimes (\ket{00}+\ket{11})_B
&=&
\\ \nonumber \ket{00}_{cc'} (\ket{00}+\ket{11})_B + \ket{01}_{cc'}
(\ket{01}+\ket{10})_B &&\\ \nonumber + i \ket{10}_{cc'} (\ket{01}-\ket{10})_B
+\ket{11}_{cc'} (\ket{00}-\ket{11})_B&& \; ,
\end{eqnarray}
which contains 2 e-bits of entanglement shared between Alice and
Bob. As entanglement does not increase under LQCC, and the
teleportation of $U$ has been done using only LQCC, we conclude
that the teleportation of a general $U$ requires at least two
e-bits.\\ Now let us proceed to show that the teleportation of an
unknown $U$ also requires the transmission of two classical bits
from Alice to Bob. The idea of the proof is to show that per
application of the CP gate Alice can transmit 2 classical bits of
information. This implies that the implementation of the CP-gate
requires 2 bits of classical communication between Alice and Bob
as otherwise we would be able to establish a super-luminal channel
between the two parties following an argument analogous to that
presented in the original teleportation paper \cite{tele}. Imagine
the following protocol. Alice encodes four messages in binary
notation as $|00\rangle,|01\rangle,|10\rangle,|11\rangle$ in two
of her control qubits. Assume that Bob holds two particles in
state $\ket{\phi^+}_B=\ket{00}+\ket{11}$, as before. The CP-gate
is applied between Alice's particle and the first of Bob's
particles (using the teleportation procedure of an unknown
operation). Depending on the state in which Alice has prepared her
two control qubits, Bob will subsequently hold one of the four
Bell states, which are mutually orthogonal. Therefore he is able
to infer Alice's message and 2 classical bits have been
transmitted. As a result, the implementation of the teleportation
of an unknown $U$ has to include the transmission of two bits of
classical information from Alice to Bob. Consider now the case
when the first of Alice's qubits is kept in a fixed state, for
instance in state $\ket{0}$. The implementation of a
controlled-Pauli operation is now equivalent to implementing a
controlled-NOT gate between Alice's second qubit and Bob's qubit
\cite{poli}. When Alice prepares the state
$\ket{+}_c=\ket{0}+\ket{1}$, the action of a controlled-NOT gate
with Bob qubit being in either state $\ket{+}_B$ or in state
$\ket{-}_B$ is given by
\begin{eqnarray}
\ket{+}_c \ket{+}_B &\longmapsto& \ket{+}_c \ket{+}_B \\ \nonumber \ket{+}_c
\ket{-}_B &\longmapsto& \ket{-}_c \ket{-}_B .
\end{eqnarray}
Therefore, this operation allows Bob to transmit one bit of
information to Alice and, as a consequence, the teleportation of
$U$ requires at least one bit of communication from Bob to Alice.
Summarizing, the physical principles of non-increase of
entanglement under LQCC and the impossibility of super-luminal
communication allow us to establish lower bounds in the resources
required for teleporting an unknown quantum operation on a qubit.
At least two e-bits of entanglement have to be consumed and, in
addition, this quantum channel has to be supplemented by a {\em
two way} classical communication channel which, in principle,
could be non-symmetric. While consistency with causality requires
two classical bits being transmitted from Alice to Bob, the lower
bound for the amount of classical information transmitted from Bob
to Alice has been found to be one bit.\\Our main result in
\cite{teleU} was to prove that the transmission of just one
classical bit from Bob to Alice is not sufficient if the protocol
is meant to work for an arbitrary $U$. We showed that each
operation $G_i$ necessarily involves a state transfer between the
remote parties and therefore, given that quantum state
teleportation can be proven to be optimal, the classical
communication cost of the remote control process is two bits in
each direction. We will analyze now what happens if the requirement
of universality is removed and characterize the sets of
transformations that can be implemented remotely without resorting
to BQST.

\section {Optimal non-trivial protocol for the implementation of an arbitrary $U$}
As explained in detail above, the basic principles establishing
the impossibility of superluminal communication and the
impossibility of increasing entanglement under LQCC allow us to
set the necessary resources for implementing a
universal remote control protocol:\\
\begin{itemize}
\item Two shared e-bits between Alice and Bob.
\item Two c-bits conveyed from Alice to Bob.
\item One c-bit conveyed from Bob to Alice.
\end {itemize}
We will now show that a protocol can be built which saturates
these bounds and achieves $50 \%$ efficiency for the remote
implementation of an arbitrary $U$. Given that the optimal
protocol consumes two classical bits from Bob to Alice, this is
the maximum probability of success if only one bit is conveyed in
that direction.\\ Our starting point can therefore be chosen a
pure state of the form
\begin{eqnarray}
\ket{\chi}_{AB} &=& \ket{\phi^+}_{AB} \otimes \ket{\phi^+}_{AB} \otimes
\ket{\psi}_b \nonumber \\&=& (\ket{00} + \ket{11})_{AB} \otimes (\ket{00} +
\ket{11})_{AB} \otimes (\alpha \ket{0} + \beta \ket{1})_b
\end{eqnarray}
where Alice and Bob share two maximally entangled states and, in
addition, Bob holds a qubit in an arbitrary state
$\ket{\psi}_b=\alpha \ket{0}_b + \beta \ket{1}_b$. In the
following we may omit at times the subscripts referring to the
parties $A$ and $B$ to make notation lighter whenever there is no
risk of confusion. The aim of the protocol is to end up with Bob
holding the transformed state $U\ket{\psi}_b$, the operation $U$
being applied only on Alice's side.

\subsection{Local actions on Bob's side}
Let us keep, for the moment, one of the shared e-bits intact. The
remaining part of the initial state can be rewritten as
\begin{equation}
\ket{\lambda}_{AB} = \alpha \ket{0}_A \ket{00}_B + \beta \ket{0}_A \ket{01}_B +
\alpha \ket{1}_A \ket{10}_B + \beta \ket{1}_A \ket{11}_B
\end{equation}
where the first qubit belongs to Alice and the other two to Bob.
We now perform a controlled-NOT operation on Bob's side using his
shared part of the e-bit as a control. After this operation, they
share the joint state
\begin{equation}
\ket{\lambda}_{AB} = (\alpha \ket{00}_{AB} + \beta \ket{11}_{AB}) \otimes
\ket{0}_B +  (\alpha \ket{11}_{AB} + \beta \ket{00}_{AB}) \otimes \ket{1}_B
\end{equation}
Bob now measures his second qubit in the computational basis. If
the result is $0$, they do noting, if it is $1$ both Alice and Bob
perform a spin flip on their qubits. As a result, Alice and Bob
now share the partially entangled state
\begin{equation}
\ket{\psi}_{AB} = \alpha \ket{00}_{AB} + \beta \ket{11}_{AB}.
\end{equation}
In this way we have managed to make the coefficients $\alpha$,
$\beta$ {\rm visible} to Alice's side or, in other words, we have
distributed the amplitudes $\alpha$ and $\beta$ onto the channel.
Note that this part of the protocol has already made use of one
e-bit. In addition, one classical bit of information has been
conveyed from Bob to Alice.
\subsection{Local actions on Alices's side}
We make now use of the extra e-bit we have kept alone so far. The
global state of the system can be written as
\begin{equation}
\ket{\lambda'}_{AB} = (\alpha \ket{00}_{AB} + \beta \ket{11}_{AB}) \otimes
(\ket{00}_{AB} + \ket{11}_{AB})
\end{equation}
Alice applies the transformation $U$ to one of her qubits. With
this, the global state reads
\begin{equation}
\ket{\psi}_{AB} = \left(\alpha\, (U\ket{0}_A) \ket{0}_B + \beta\,
(U \ket{1}_A) \ket{1}_B \right) \otimes (\ket{00}_{AB} +
\ket{11}_{AB}) \label{ab}
\end{equation}
The remaining part of the protocols mimics quantum state teleportation with
Alice performing a Bell measurement on her side. This procedure makes use of
the extra e-bit and involves the transmission of two classical bits of
information from Alice to Bob. We will see in the following that as a result of
this protocol, Bob ends up holding a two-qubit state of the form:
\begin{equation}
(\alpha \, U\ket{0} + \beta \, U\ket{1})\otimes \ket{0} + (\alpha \, U\ket{0} -
\beta \, U\ket{1}) \otimes \ket{1}=U(\ket{\psi}_b)\otimes\ket{0}+U(\sigma_z
\ket{\psi}_b) \otimes \ket{1}
\end{equation}
A final projective measurement on Bob's side yields the correct transformed
state with $50 \%$ probability, the maximum allowed when the transformation $U$
is completely arbitrary and only one bit is conveyed from Alice to Bob.
\subsubsection{Detailed steps}
For our purposes, it suffices to parametrize the transformation $U$ as a
generic unimodular matrix, i.e. an arbitrary rotation on a qubit of the form
\begin{equation}
    U = \left(
    \begin{array}{cc}
    a & b\\
    -b^* &  a^*
    \end{array}
    \right)
    \label{uu}
\end{equation}
where the coefficients $a$ and $b$ obey the unimodular constraint
$|a|^2 + |b|^2=1$ 
Using the Bell basis $(\ket{\phi^{\pm}}_A=\ket{00}_A \pm
\ket{11}_A,\ket{\psi^{\pm})}_A=\ket{01}_A \pm \ket{10}_A$ for
Alice's qubits, we can rewrite the joint state given by eq.
(\ref{ab}) as follows
\bea
\ket{\lambda'}_{AB} &=&
\ket{\phi^+}_{A} \otimes (\alpha \ket{0} U\ket{0} + \beta \ket{1}
U\ket{1}) \nonumber \\ &+&\ket{\phi^-}\otimes (\id \otimes
\sigma_z)(\alpha \ket{0} U\ket{0} + \beta \ket{1} U\ket{1})
\nonumber \\ &+& \ket{\psi^+}\otimes(\id \otimes \sigma_x)(\alpha
\ket{0} U\ket{0} + \beta \ket{1} U\ket{1}) \nonumber \\&+&
\ket{\psi^-}\otimes(\id \otimes \sigma_x\sigma_z)(\alpha \ket{0}
U\ket{0} + \beta \ket{1} U\ket{1})
\eea
Alice now performs a Bell
measurement on her two qubits and informs of her results to Bob
using a classical channel. Accordingly to Alice's measurement
outcomes, Bob performs on his second qubit the same operations as
the corresponding to the protocol of quantum state teleportation.
As a result of this procedure, he always ends up holding the
following two qubit (pure) state $$ \alpha \ket{0} (a \ket{0} + b
\ket{1}) + \beta \ket{1} (-b^* \ket{0} + a^* \ket{1}) $$, which
after a local Hadamard transformation on the first qubit reads $$
\ket{0} \otimes (\alpha\, U\ket{0} +
 \beta\,
 U\ket{1}) + \ket{1} \otimes (\alpha\, U\ket{0} - \beta\,
 U\ket{1})
$$ A final projective measurement on the first qubit leaves Bob holding the
correct transformed state by $U$ whenever the measurement outcome is $0$.
However, in the case that the local measurement throws the outcome $1$, Bob
would hold the wrong state $\alpha\, U\ket{0} - \beta\, U\ket{1}$ and, provided
that $U$ is completely arbitrary, he cannot restore this state to correct form.
As a result, the protocol is successful in $50 \%$ of the cases. Note that this
is the maximum efficiency we can expect when only one bit is conveyed from Bob
to Alice. It is a remarkable fact, and a direct consequence of the linearity of
quantum mechanics \cite{teleU}, that no protocol different from bi-directional
quantum state teleportation can achieve the remote implementation of any
arbitrary operation on a qubit. But, are there sets of transformations for
which is possible for Bob to restore the final state to the correct form
$\alpha\, U\ket{0} + \beta\, U\ket{1}$?

\section{Restricted set of operations}
As discussed in detail in the previous section, with probability
$50\%$ Bob is left holding the wrong transformed state
\begin{equation}
  \alpha \, U \ket{0} - \beta \, U \ket{1} = U \sigma_z \ket{\phi}_B
\end{equation}
Given that the transformation $U$ given by Eq.(\ref{uu}) is completely unknown
to him, no subsequent local action can yield the correct transformed state $U|\phi\rangle_B$
for
every $U$. However, it is clear from the above expression that there are cases
where implementing a universal correction operation $V$ is possible. Formally,
we are seeking for an operator $V$ such that
\begin{equation}
  V  U \sigma_z \ket{\phi}_B = e^{i \delta}\, U \ket{\phi}_B
\end{equation}
for any $\ket{\phi}_B$, $\gamma$ being a real parameter. Therefore, the
following operator identity must hold
\begin{equation}
  V U = e^{i \delta} \, U \sigma_z. \label{corr}
\end{equation}
We can immediately identify a set of transformations that can be remotely
implemented. If we set $V=\sigma_z$, the two possible uni-modular solutions to
Eq.(\ref{corr}) are given by (up to a local change of basis)\cite{basis}:
\begin{equation}
    U_c = \left(
    \begin{array}{cc}
    a & 0\\
    0 &  a^*
    \end{array}
    \right) = e^{i\phi\sigma_z},
    \label{uu1}
\end{equation}
with $a=e^{i\phi}$ that is, the set of operations that commute with
$\sigma_z$, or transformations of the form
\begin{equation}
    U_a = \left(
    \begin{array}{cc}
    0 & b\\
    -b^* &  0
    \end{array}
    \right),
    \label{uu2}
\end{equation}
with $b=e^{i\phi}$ which anticommute with $\sigma_z$, i.e., are linear combinations
of the Pauli operators $\sigma_x$ and $\sigma_y$. Any operation within this family can be
teleported with $100 \%$ efficiency using a protocol which employs less
resources than BQST. We can physically interpret the set of allowed transformations as
\begin{itemize}
\item Arbitrary rotations around the z-axis.
\item Rotations by $\pi$ around any axis lying within the equatorial plane.
\end{itemize}
We will illustrate in the next section, using the Bloch sphere representation
for qubits, how it can be easily visualized why a universal correction by means
of the application of the operator $\sigma_z$ is possible is these cases.\\
There is still a question that remains to be addressed. Are the sets of
operations we have just described the only ones that can be implemented
remotely by {\rm non-trivial} means? We will postpone the issue of uniqueness
till Section VI.
\section{Geometrical interpretation}
The aim of this section is just
to provide an intuitive geometrical picture in order to visualize
which transformations can be implemented remotely by
non-trivial means and illustrate the role of the final restoration
step on Bob's side. Let us consider first a very simple scenario
in which Bob is holding a qubit state lying in the equatorial
plane of the Bloch sphere,
\begin{equation}
\ket{\phi}_b = \alpha \ket{0} + \beta \ket{1} = \frac{1}{\sqrt 2} (\ket{0} +
e^{i \zeta} \ket{1}).
\end{equation}
Imagine now that the transformation we want to implement remotely is just a
spin flip, i.e., $U=\sigma_x$ (Obviously Alice does not know this!). In this
case, given that the Pauli operator anticommute with $\sigma_z$, the protocol described in the
previous section will result in Bob to hold the correct transformed state $\sigma_x
\ket{\phi}_B$. If Bob follows the prescribed rules, prior to the final
correction step with $50 \%$ probability he holds the correct transformed state
and with $50 \%$ probability he holds the erroneous state
\begin{equation}
\ket{\psi}_{b,W} = \alpha U \ket{0} - \beta U \ket{1}= U (\frac{1}{\sqrt 2}
(\ket{0} - e^{i \zeta} \ket{1}).
\end{equation}
Therefore, we can also consider the wrong transformed state as the transformed
by $U$ of the qubit state $\ket{\bar \psi}_b=\sigma_z \ket{\psi}_b$. Which
state Bob ends up holding depends on certain measurement outcome and therefore
he knows whether a subsequent correction step is necessary or not. The relative
position of the Bloch vectors representing the initial states $\ket{\psi}_B$
and $\ket{\bar \psi}_B$ and their transformed vectors by $U$ are shown in
figure 1. In this case, states $\ket{\psi}_B$ and $\ket{\bar \psi}_B$ are
orthogonal and their associated Bloch vectors lie opposite in the equatorial
plane of the Bloch sphere. The action of $U$ preserves the relative orientation
and the Bloch vectors associated with the transformed states by U, dashed lines
in the figure, are opposite as well. The key point is that a subsequent
application of the operation $\sigma_z$ onto the wrong transformed state just
flips its Bloch vector and yields the correct state. These considerations may
sound rather trivial but it is all we need to intuitively understand how the
protocol works in the general case.
\begin{figure}[tbp]
\epsfxsize6.5cm
 \centerline{\epsfbox{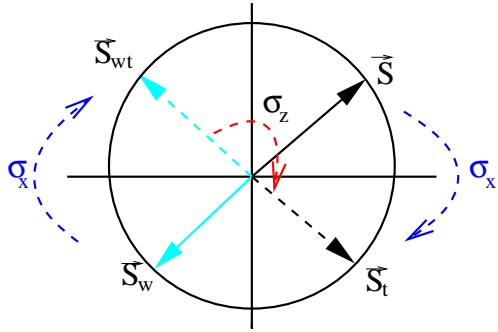}}
 \vspace*{0.2 cm}
\caption{Geometrical interpretation of the restoration to the correct
transformed state when the transformation $U$ belongs to a restricted set. See
the text for details.}
\end{figure}
Imagine now that the transformations $U$ is not simply a Pauli operator but a
transformation of the general form given by eq.(\ref{uu}). Assuming that Bobs
state lies initially in the equatorial plane, as before, the corresponding
Bloch vector of the transformed state by $U$ does not longer lie onto the
equator of the Bloch sphere and has in general a non-zero z-component
$S_z= |\alpha|^2 - |\beta|^2$ (we have defined $S_i=tr\rho\sigma_i$)
However, this component is equal to zero if the
transformation $U$ either commutes or anticommutes with $\sigma_z$ (operations
of the form $U_c$ or $U_a$). In this case, we recover the situation discussed
before. The transformed Bloch vectors lie opposite along some direction
contained in the equatorial plane and a final step via the application of
$\sigma_z$ restores the wrong transformed state to the correct one.\\ What
happens in general? The easiest way to analyze the general case, where Bob
holds an arbitrary qubit state, is to parametrize it as a generic spinor and
split the representation in terms of the associated Bloch vectors into two
components as follows
\begin{equation}
    \rho_b = \ket{\psi} \bra{\psi}=\frac{1}{2} ( \id + S_{x}\sigma_{x} + S_{y}\sigma_{y}
    + S_z\sigma_z). \label{blo1}
\end{equation}
Analogously, the wrong transformed state can be thought of as obtained from $U$
acting upon the state
\begin{equation}
    \bar\rho_b = \ket{\bar \psi} \bra{\bar \psi}=\frac{1}{2} ( \id -
    S_{x}\sigma_{x} - S_{y}\sigma_{y} + S_z\sigma_z),\label{blo2}
\end{equation}
so we can write the erroneous transformed state as
\begin{equation}
    U \bar \rho_b U^\dag =\frac{1}{2} ( \id - U(S_{x}\sigma_{x} + S_{y}\sigma_{y}) U ^\dag+ U
    S_z\sigma_z U^\dag).
\end{equation}
Consider the case where the transformation $U$ commutes with the action of
$\sigma_z$. When Bob applies the final correction step, the transformed state
reads
\begin{eqnarray}
    \sigma_z U \bar \rho_b U^\dag \sigma_z&=&\frac{1}{2} ( \id - \sigma_z U
    (S_{x}\sigma_{x} + S_{y}\sigma_{y}) U ^\dag \sigma_z+ \sigma_z U S_z\sigma_z U^\dag
    \sigma_z) \nonumber \\ &=&
    \frac{1}{2} ( \id + U (S_{x}\sigma_{x} + S_{y}\sigma_{y}) U^\dag +
    U S_z\sigma_z U^\dag) = U \rho_B U^\dag,
\end{eqnarray}
where we have taken into account that Pauli operators anti-commute among
themselves and that $\sigma^2=\id$. A similar argument holds for the case where
$U$ anti-commutes with $\sigma_z$. Resuming our geometrical picture, in the
general case the corresponding Bloch vectors associated to the states
$\ket{\psi}_b$ and $\ket{\bar \psi}_b$ have the same z-component while the
corresponding projections onto the equatorial plane lie opposite. Therefore,
under the action of a transformation $U$ which either commutes or anticommutes
with $\sigma_z$, we recover the situation discussed at the beginning of this
section and a final correction by means of applying the operation $\sigma_z$
restores the correct transformed state.

\section{Characterization of sets that allow remote implementation without bidirectional
state teleportation}
So far we have identified two sets of transformations that can be
implemented remotely without resorting to BQST (bidirectional
state teleportation). However, the procedure by which they have
been identified does not allow to draw any conclusion as far as
their uniqueness is concerned. This is the aim of this section. To
do this we first establish necessary conditions for avoiding BQST
and then we show the uniqueness of the two sets of
transformations.

\subsection{Necessary conditions for avoiding BQST}
Let the set of operators that can be remotely implemented on Bob's
qubit be labeled  as ${\cal U}$.  We know \cite{teleU} that if
${\cal U}$ is the full set of unimodular operations on a qubit,
the protocol necessarily teleports the state of Bob's qubit to
Alice, that is, every state undergoes BQST. In contrast, in the
protocol described in Section III, where ${\cal U}$ contains
operators of the form Eqs.\ (\ref{uu1}) and (\ref{uu2}), it is
easy to show that only the two orthogonal states $\ket{0}$ and
$\ket{1}$ undergo BQST. In this section we examine the
relationship between the size of the set ${\cal U}$ and the number
of states that undergo BQST. From this we show that if Bob is
restricted to sending 1 c-bit to Alice, then the set ${\cal U}$
comprises two particular subsets.

\subsubsection{Subsets of operators}

In \cite{teleU} we showed that the if the teleported operation $U$
is arbitrary, that is ${\cal U}$ is the full set of unimodular
operations, the final state of the ancilla is independent of $U$.
However, here the set of operators ${\cal}$ is restricted and so
the final state of the ancilla may depend on which operation is
teleported.  Hence we reexpress the operation of the black box as
(cf \eqr{doble})
\begin{equation}
   G_2 U_n G_1 (\ket{\chi}_{aAB}\ket{\psi}_{b}) =
   \ket{\Phi(\chi,U_n)}_{aAB}U_n\ket{\psi}_{b}
\end{equation}
for $U_n\in{\cal U}$. We know that the final state
$\ket{\Phi(\chi,U_n)}_{aAB}$ is independent of $\ket{\psi}_{b}$ by
the same arguments presented in our previous work \cite{teleU}.

Consider action of the gate for a linear combination of operators
$U=\sum_n c_n U_n$ where $U, U_n\in {\cal U}$:
\begin{eqnarray*}
   G_2 U_n G_1 \ket{\chi}_{aAB}\ket{\psi}_{b}
   &=&  \sum_n  c_n  G_2 U_n G_1 \ket{\chi}_{aAB}\ket{\psi}_{b} \\
   &=&  \sum_n  c_n  \ket{\Phi(\chi,U_n)}_{aAB}U_n\ket{\psi}_{b}
\end{eqnarray*}
which equals $\ket{\Phi(\chi,U)}_{aAB}U\ket{\psi}_{b}$ only if
$\ket{\Phi(\chi,U_n)}_{aAB}$ is independent of $U_n$.  In other
words, linearly dependent operators share the same final state.
This final state may depend on set of linearly dependent control
operators, however. Indeed, we subdivide the set ${\cal U}$ into
subsets
\[
   {\cal U}={\cal U}^{(1)}\cup {\cal U}^{(2)}\cup\ldots
\]
which leave the state of the ancilla in the same final state:
\[
    \ket{\Phi(\chi,U_i^{(n)})}
    =\ket{\Phi(\chi,U_j^{(n)})}
    =\ket{\Phi^{(n)}(\chi)}
\]
where $U_i^{(n)}\in {\cal U}^{(n)}$.  [We use the superscript ``$(n)$'' to
label a subset and its elements.] It follows that the subsets ${\cal U}^{(n)}$
are linearly independent in the sense that an operator in one subset cannot be
written as a linear combination of operators from other sets. Also, the subsets
are clearly disjoint as each operator $U\in{\cal U}$ belongs to one and only
one subset ${\cal U}^{(n)}$. Since there are a maximum of 4 linearly
independent operators on the 2 dimensional state space, there are thus a
maximum of 4 subsets ${\cal U}^{(n)}\subset{\cal U}$.

\subsubsection{Special case $G_1=\id$}

It is interesting to consider the special case where $G_1=\id$. We
now show that for this case there are a maximum of 4 operators
which can be teleported. Consider two operators, $U_1^{(n)}$ and
$U_2^{(m)}$ and choose an orthogonal pair of states
$\ip{\psi|\psi_\perp}=0$ such that
\begin{eqnarray}
  U_1^{(n)}\ket{\psi}&=&\ket{\phi}\ ,
    \label{Upsi}\\
  U_2^{(m)}\ket{\psi_\perp}&=&\ket{\phi^\prime}
    \label{Upsi_perp}
\end{eqnarray}
where $\ip{\phi|\phi^\prime}\ne 0$ and $U_i^{(k)}\in{\cal
U}^{(k)}$.  The fact that this is possible is proved in the
Appendix. Thus, we can write
\begin{eqnarray}
  U_1^{(n)}\ket{\chi}_{aAB}\ket{\psi}_{b}
        &=&G^\dagger_2\ket{\Phi^{(n)}(\chi)}_{aAB}
                                        U_1^{(n)}\ket{\psi}_{b}
          \nonumber\\
        &=&G^\dagger_2\ket{\Phi^{(n)}(\chi)}_{aAB}\ket{\phi}_{b}\ ,
        \label{G1psi}\\
  U_2^{(m)}\ket{\chi}_{aAB}\ket{\psi_\perp}_{b}
        &=&G^\dagger_2\ket{\Phi^{(m)}(\chi)}_{aAB}
                                        U_1^{(m)}\ket{\psi_\perp}_{b}
          \nonumber\\
        &=&G^\dagger_2\ket{\Phi^{(m)}(\chi)}_{aAB}\ket{\phi^\prime}_{b}\ .
        \label{G1psi_perp}
\end{eqnarray}
The inner product of the left-hand sides of Eqs.\ (\ref{G1psi}) and
(\ref{G1psi_perp}) is zero and so
\begin{equation}
   0=\ip{\Phi^{(n)}(\chi)|\Phi^{(m)}(\chi)}_{aAB}\
      \ip{\phi|\phi^\prime}_{b}\ .
   \label{innerprod}
\end{equation}
But if $n=m$, then
$\ip{\Phi^{(n)}(\chi)|\Phi^{(m)}(\chi)}_{aAB}=1$ and
\eqr{innerprod} cannot be satisfied.  We conclude that each subset
contains only one operator.

Also, if $n\ne m$ (i.e. different subsets) then \eqr{innerprod}
implies $\ip{\Phi^{(n)}(\chi)|\Phi^{(m)}(\chi)}_{aAB}=0$ and so
the final ancilla states are orthogonal. The number of operators
able to be teleported therefore depends on the dimension of the
ancilla state space.  Provided this can be made large enough,
there will be a maximum of 4 operators able to be teleported with
$G_1=\id$ (because there are a maximum of 4 linearly-independent
subsets).

The fact that the final states of the ancilla are orthogonal for
different operators means that the operators themselves are
orthogonal. Imagine that Alice has a son called Bobby in her lab.
She teleports the operator to Bobby and together they examine the
state of the final state of their (local) ancilla. From this they
can determine which operator Alice teleported.  Alice can
communicate this to Bob using 2 classical bits of information, and
Bob can then carry out locally the corresponding operation on his
qubit.

Hence the special case $G_1=\id$ leads to a trivial classical
remote control scenario.  For the remainder of this paper we only
consider the case where $G_1\ne\id$.

\subsubsection{Conditions for the BQST of a state}

We now look at a sufficient condition on the set ${\cal U}$ for
the BQST of a state. This will give us a necessary condition for
avoiding BQST for a set of states.  Choose $U^{(n)}\in {\cal
U}^{(n)}$ and let
\begin{equation}
   U^{(n)}\ket{\psi}=\ket{\phi}\ .
   \label{def_phi}
\end{equation}
Thus we have
\begin{equation}
   G_2 U^{(n)} G_1 \ket{\chi}_{aAB}\ket{\psi}_{b}
   = \ket{\Phi^{(n)}(\chi)}_{aAB} U^{(n)}\ket{\psi}_{b}
   = \ket{\Phi^{(n)}(\chi)}_{aAB} \ket{\phi}_{b}
   \label{GUG}
\end{equation}
and so
\begin{equation}
   G_1 \ket{\chi}_{aAB}\ket{\psi}_{b}
   = [U^{(n)}]^\dagger G_2^\dagger\ket{\Phi^{(n)}(\chi)}_{aAB} \ket{\phi}_{b}\ .
   \label{eigen1}
\end{equation}
Next we construct the unimodular operator $Q(\alpha,\ket{\xi})$ as follows
\begin{equation}
   Q(\alpha,\ket{\xi}) \equiv e^{i\alpha}\ket{\xi}\bra{\xi}
                    + e^{-i\alpha}(\id - \ket{\xi}\bra{\xi})
   \label{def_Q}
\end{equation}
for $\alpha\neq 0,\pi, 2\pi, \ldots$ and arbitrary (normalised) state
$\ket{\xi}$.  This operator has the property that
\[
   Q(\alpha,\ket{\phi})U^{(n)}=U^{(n)}Q(\alpha,\ket{\psi})\ .
\]
If $U^{(n)}Q(\alpha,\ket{\psi})\in {\cal U}^{(n)}$ then we can
replace $U^{(n)}$ in \eqr{GUG} with $U^{(n)}Q(\alpha,\ket{\psi})$
and obtain from \eqr{eigen1}
\begin{eqnarray}
 \non
 Q(\alpha,\ket{\phi}_{a}) G_1 \ket{\chi}_{aAB}\ket{\psi}_{b}
   &=& [U^{(n)}]^\dagger G_2^\dagger\ket{\Phi^{(n)}(\chi)}_{aAB}
       U^{(n)} Q(\alpha,\ket{\psi}_{b})\ket{\psi}_{b}\\
   &=& e^{i\alpha} [U^{(n)}]^\dagger G_2^\dagger
       \ket{\Phi^{(n)}(\chi)}_{aAB} \ket{\phi}_{b}\ .
   \label{eigen2}
\end{eqnarray}
Comparing \eqr{eigen1} with \eqr{eigen2} shows that $G_1
\ket{\chi}_{aAB}\ket{\psi}_{b}$ is an eigenstate of
$Q(\alpha,\ket{\psi}_{a})$, i.e.
\[
   G_1 \ket{\chi}_{aAB}\ket{\psi}_{b} = \ket{\psi}_{a}\otimes\ldots
\]
or, in other words, that the state of Bob's qubit is necessarily teleported to
Alice by the operation of $G_1$.  Note that if $U^{(n)}Q(\alpha,\ket{\psi})$
belongs to a different subset, say ${\cal U}^{(m)}$ with $m\ne n$, then instead
of \eqr{eigen2} we get
\begin{eqnarray*}
 Q(\alpha,\ket{\phi}_{a}) G_1 \ket{\chi}_{aAB}\ket{\psi}_{b}
   &=& [U^{(n)}]^\dagger G_2^\dagger\ket{\Phi^{(m)}(\chi)}_{aAB}
       U^{(n)} Q(\alpha,\ket{\psi}_{b})\ket{\psi}_{b}\\
   &=& e^{i\alpha} [U^{(n)}]^\dagger G_2^\dagger
       \ket{\Phi^{(m)}(\chi)}_{aAB} \ket{\phi}_{b}\\
   &\ne& e^{i\alpha} [U^{(n)}]^\dagger G_2^\dagger
       \ket{\Phi^{(n)}(\chi)}_{aAB} \ket{\phi}_{b}\ .
\end{eqnarray*}
and so the state of Bob's qubit is {\em not} teleported by $G_1$
to the qubit operated on by $U^{(n)}$.  Hence we can state a
sufficient condition for BQST as follows: BQST occurs for a state
$\ket{\psi}$ when at least one value of $\alpha\neq
0,\pi,2\pi,\ldots$ can be found such that
$U^{(n)}Q(\alpha,\ket{\psi})\in{\cal U}^{(n)}$ for at least one
operator $U^{(n)}\in{\cal U}^{(n)}$ for any ${\cal
U}^{(n)}\subset{\cal U}$.

Consider, for the moment, the case where we insist that {\em none}
of the states $\ket{\psi}$ undergo BQST.  This requires that
$U^{(n)}Q(\alpha,\ket{\psi})\notin {\cal U}^{(n)}$ for all
$\ket{\psi}$, all $U^{(n)}\in {\cal U}^{(n)}$, all ${\cal
U}^{(n)}\in {\cal U}$ and all $\alpha\neq 0,\pi,2\pi,\ldots$.  The
set of $Q(\alpha,\ket{\psi})$ for all $\ket{\psi}$ and all
$\alpha\neq 0,\pi,2\pi,\ldots$ is the set of all unimodular
operators minus the the set of operators which are proportional to
the identity. Assume for the moment that ${\cal U}^{(n)}$ contains
the two operators $U_1^{(n)}$, $U_2^{(n)}$ where $U_1^{(n)}\ne
e^{i\theta}U_2^{(n)}$ for any real $\theta$. We can set
$Q(\alpha,\ket{\psi})=[U_1^{(n)}]^\dagger U_2$ for an appropriate
choice of $\ket{\psi}$ and $\alpha$, and so $U_1^{(n)}
Q(\alpha,\ket{\psi})=U_2^{(n)}\in {\cal U}^{(n)}$. This means that
the state $\ket{\psi}$ would be BQST contradicting our starting
point. Clearly if no states are to undergo BQST then each subset
${\cal U}^{(n)}$ cannot contain more than one operator (up to an
imaginary phase factor).  Hence, for the case where no states are
BQST,  ${\cal U}$ contains at most 4 linearly independent
operators. We note that if the 4 operators are orthogonal (i.e.
related to the identity operator and 4 Pauli operators by a fixed
transformation) Alice may distinguish between them using local
means and thus send 2 classical bits to Bob who could then
implement locally the appropriate operation on his qubit.  The
general case, however, would require either more measurements by
Alice to determine the operator, or a more sophisticated channel
between Alice and Bob (i.e. with shared ebits etc.)

Returning to the more general case, one can see from \eqr{def_Q}
that $Q(\alpha,\ket{\psi})=Q(-\alpha,\ket{\psi_\bot})$ where
$\ip{\psi_\bot|\psi}=0$ and so if $\ket{\psi}$ undergoes BQST then
so to are the states orthogonal to $\ket{\psi}$. Non-trivial
remote control therefore necessarily incurs BQST for at least one
pair of orthogonal states.  Bob can communicate 1 classical bit to
Alice by preparing his qubit in one of these orthogonal states and
stopping the protocol after $G_1$. The scheme we are most
interested in is where Bob sends exactly 1 classical bit of
information to Alice.  Henceforth we only consider the case where
exactly one pair of orthogonal states undergo BQST with all other
states avoiding BQST.

\subsubsection{BQST of a single pair of states}

For brevity we take the pair of orthogonal states that are BQST to
be the computational basis states: $\ket{0}$, $\ket{1}$.  (It is
straight forward to generalize our analysis to an arbitrary pair.)
All other states,
\[
   \ket{\psi'} = a\ket{0} + b \ket{1}
\]
for $a$, $b\ne 0,1$, do not undergo BQST.  We can write this as
\[
    U_i^{(n)}Q(\alpha,\ket{\psi'}) \notin {\cal U}^{(n)}
\]
or, equivalently,
\[
    Q(\alpha,\ket{\psi'}) \notin [U_i^{(n)}]^\dagger{\cal U}^{(n)}
\]
for all $\alpha\ne 0, \pi, 2\pi, \ldots$, all $\ket{\psi'}\ne
\ket{0}, \ket{1}$, all $U_i^{(n)}\in{\cal U}^{(n)}$ and all
subsets ${\cal U}^{(n)}\subset {\cal U}$. The set of operators
$\{Q(\alpha,\ket{\psi'})\}$ here is the set of all unimodular
operators {\em not} diagonalized by $\ket{0}, \ket{1}$. Hence each
set $[U_i^{(n)}]^\dagger{\cal U}^{(n)}$ contains operators which
{\em are} diagonalized by $\ket{0}, \ket{1}$. Thus all elements of
each subset ${\cal U}^{(n)}\subset {\cal U}$ have the form
\begin{eqnarray}
   U_\beta^{(n)}
      &=& U_0^{(n)}(e^{i\beta}\ket{0}\bra{0}+e^{-i\beta}\ket{1}\bra{1})
          \non \\
      &=& U_0^{(n)} Q(\beta,\ket{0}) \non \\
      &=& U_0^{(n)} e^{i\beta\sigma_z}\ .
   \label{diag_form}
\end{eqnarray} If the subsets ${\cal U}^{(n)}$ are the largest possible (i.e. ${\cal
U}^{(n)}$ contains the operators $U_\beta^{(n)}$ for all $\beta$) then there
are a maximum of 2 subsets ${\cal U}^{(n)}\subset {\cal U}$.  To see this
consider an arbitrary, unimodular, linear combination of the elements of 2
subsets ${\cal U}^{(1)}$ and ${\cal U}^{(2)}$:
\[
   U = x U_0^{(1)}e^{i\beta\sigma_z} + y U_0^{(2)}e^{i\gamma\sigma_z}
\]
where $x$ and $y$ are real numbers.   We can write this as
\[
  [U_0^{(1)}]^\dagger U = x e^{i\beta\sigma_z} +
                y [U_0^{(1)}]^\dagger U_0^{(2)}e^{i\gamma\sigma_z}
\]
or, in matrix form, as \begin{eqnarray*}
  \left[\begin{array}{cc}
    c       & d \\
    -d^\ast & c^\ast
  \end{array}\right]
  =
  x \left[\begin{array}{cc}
     e^{i\beta} & 0 \\
     0          & e^{-i\beta}
    \end{array}\right]
  + y  \left[\begin{array}{cc}
        a e^{i\gamma}       & b e^{-i\gamma} \\
        -b^\ast e^{i\gamma} & a^\ast e^{-i\gamma}
       \end{array}\right]
\end{eqnarray*}
where
\begin{eqnarray*}
  [U_0^{(1)}]^\dagger U =
  \left[\begin{array}{cc}
    c       & d \\
    -d^\ast & c^\ast
  \end{array}\right]
\end{eqnarray*}
\begin{eqnarray*}
   [U_0^{(1)}]^\dagger U_0^{(2)} =
  \left[\begin{array}{cc}
    a       & b \\
    -b^\ast & a^\ast
    \end{array}\right]\ .
\end{eqnarray*}
Clearly, $y e^{-i\gamma} = d/b$ and $x e^{i\beta}=c-yae^{i\gamma}$ which
can be solved for real values of $x$, $y$, $\beta$ and $\gamma$ for arbitrary
$c$ and $d$ satisfying $|c|^2+|d|^2=1$. This shows that every unimodular
operator $U'=[U_0^{(1)}]^\dagger U$, and hence every unimodular operator
$U=U_0^{(1)}U'$, can be written in terms of a linear combination of operators
in the subsets ${\cal U}^{(1)}$ and ${\cal U}^{(2)}$. These two subsets are,
therefore, the only linearly independent subsets.

We note that restricting the number of subsets to 1 and choosing either
$U_0^{(1)}=\id$ or $U_0^{(1)}=i\sigma_y$ in \eqr{diag_form} corresponds to the
situation in the $1-1-1$ protocol. On the other hand choosing $U_0^{(1)}=\id$
and $U_0^{(2)}=i\sigma_y$ corresponds to the situation in the $2-2-1$ protocol.

To sum up this subsection:  to avoid BQST for all states, the set
of control operators must be restricted to a set of a maximum of 4
linearly independent operators; if one state undergoes BQST then
so are the states orthogonal to it; if Bob is restricted to
sending 1 classical bit to Alice then only 1 pair of orthogonal
states can undergo BQST and the set of control operations ${\cal
U}$ can be divided into a maximum of 4 subsets ${\cal
U}^{(n)}\subset {\cal U}$ whose elements have the form
\eqr{diag_form}; if the subsets ${\cal U}^{(n)}$ in
\eqr{diag_form} contain operators $U_\beta^{(n)}$ for all $\beta$
then only 2 subsets are possible.

Finally, we note that these conditions on the set ${\cal U}$ of controlled
operators are {\em necessary} for avoiding the BQST of various states. They are
not {\em sufficient} conditions because $G_1$ and $G_2$ can be chosen to
perform BQST for all states, irrespective of the restrictions on ${\cal U}$.

\subsection{Full characterization of classes of operators allowing for non-trivial
remote implementation.}
In this subsection we now wish to complete the characterization of the classes
of state that can be implemented without BQST. We will show that a protocol
that consumes 2 shared ebits + 2 bit of classical communication from
A$\rightarrow$B + 1 bit of classical communication from B$\rightarrow$A (221)
for teleportation of unitary operations is only possible when the operations
are drawn from the following two sets:

\begin{eqnarray}
    \mbox{Set A}:&& \left[ \left(\begin{array}{cc} e^{i\phi} & 0 \\ 0 & e^{-i\phi} \end{array}\right),
    \phi\in \mathbb{R} \right] \label{set1}\\
    \mbox{Set B}:&& \left[ \left(\begin{array}{cc} a & b \\ -b^* & a^* \end{array}\right)
    \left(\begin{array}{cc} e^{i\phi} & 0 \\ 0 & e^{-i\phi} \end{array}\right),
    \phi\in \mathbb{R} \right] \label{set2}
\end{eqnarray}
under the constraint that either $|a|=1$ (trivial) or $|b|=1$. Any other
choices will require more resources \footnote{Of course we have in addition the
freedom of choice of basis, ie we can change all the above sets jointly by a
fixed basis change, but that is a trivial freedom}. Together with the results
from the previous section this then concludes our characterization of those
operations that allow for non-trivial remote implementation.

As outlined in section II and used throughout this paper the most general
of any possible protocol is given by
\begin{equation}
    G_2 U G_1 |\chi\rangle_{A}|\psi\rangle_B = |\bar\chi\rangle
    (U|\psi\rangle)_{B} \; . \label{start}
\end{equation}
where without loss of generality the state $|\chi\rangle$ is a tensor product
state. For whatever form of $G_1$ we can always write Eq. \ref{start} as
\begin{eqnarray}
    G_2UG_1 |\chi\rangle_A|\psi\rangle_B &=& G_2((U|0\rangle_A)|\Phi_0\rangle_{AB} +
    (U|1\rangle_A)|\Phi_1\rangle_{AB}) = |\bar\chi\rangle_{AB}
    (U|\psi\rangle)_{B}\; .
    \label{eq45}
\end{eqnarray}
We can always write $|\psi\rangle=\alpha|0\rangle+\beta|1\rangle$ and note that
$|\bar\chi\rangle$ is independent from both $\phi$ and $|\psi\rangle$, but may
of course depend on $a$ and $b$. From normalization we have
$\langle\Phi_0|\Phi_0\rangle+\langle\Phi_1|\Phi_1\rangle=1$. If we now evaluate eq. \ref{eq45}
for two unitaries $U_1,U_2$ from the above sets Eqs. \ref{set1}-\ref{set2} we
can obtain the following scalar product
\begin{eqnarray}
    \sum_{ij} \langle i|U_2^{\dagger}U_1|j\rangle\langle\Phi_i|\Phi_j\rangle
    = (|\alpha|^2 \langle 0|U_2^{\dagger}U_1|0\rangle+
    |\beta|^2 \langle 1|U_2^{\dagger}U_1|1\rangle+
    \alpha\beta^* \langle 1|U_2^{\dagger}U_1|0\rangle+
    \alpha^*\beta \langle 0|U_2^{\dagger}U_1|1\rangle)\langle\bar\chi_2|\bar\chi_1\rangle
    \; .
\end{eqnarray}
The proof proceeds in essentially two steps. First we will demonstrate that in
the protocol the operation $G_1$ will generally create an entangled state between
the qubit $U$ is acting upon and the rest of the systems. Up
to local rotations any entangled state is of the form
$r|00\rangle+s|11\rangle$. In the basis where the entangled state can be
written like this we will then show, that when $U$ acts on it we can only find a
$G_2$ that recovers $U|\psi\rangle$ if either $|a|=1$ or $|b|=1$. This then
concludes the proof.\\

i) {\em Assume that there is no entanglement generated by $G_1$.}\\
Given that the set of transformations that we want to teleport is non-trivial,
ie they are generally non-orthogonal, the transformation $G_1$ has to be non-trivial. This
implies in particular that a strategy of distinguishing the unitaries is not
possible. Therefore we cannot have the situation that
$|\Phi_0\rangle=|\Phi_1\rangle=|\psi\rangle$ for all unitaries $U$. \footnote{This
remark is relevant because for the case of four orthogonal transformations the
following argument does not hold, because we assume that $|\chi\rangle$ is
independent of $U$ which only needs to hold when one wishes to teleport
non-orthogonal transformations!}

Now let us assume that $G_1$ does not generate an entangled state which requires that
\begin{equation}
    |\Phi_0\rangle = \frac{x'}{y'}|\Phi_1\rangle
\end{equation}
Under this assumption we will now demonstrate that then $x'/y' = \alpha/\beta$.
To this end let us make the special choice $U_2={\bf 1}$ which simplifies the analysis
and is sufficient to generate the desired result. Then we have
\begin{eqnarray}
    \langle 0|U_1|0\rangle |x'|^2 + \langle 1|U_1|1\rangle |y'|^2
    + \langle 0|U_1|1\rangle (x')^*y' + \langle 1|U_1|0\rangle x'(y')^*
    &=& \\
    (|\alpha|^2 \langle 0|U_1|0\rangle +
    |\beta|^2 \langle 1|U_1|1\rangle+
    \alpha\beta^* \langle 1|U_1|0\rangle && \!\!\!\!\!\!+
    \alpha^*\beta \langle 0|U_1|1\rangle) g \nonumber
\end{eqnarray}
where $g$ depends on whether $U_1$ is chosen from set A ($g=1$) or set B ($g$
to be determined in a moment). Therefore
\begin{eqnarray}
    \hspace*{-1.cm}\langle 0|U_1|0\rangle (|x'|^2-g|\alpha|^2) + \langle 1|U_1|1\rangle
    (|y'|^2-g|\beta|^2)
    + \langle 0|U_1|1\rangle ((x')^*y'-g\alpha^*\beta) + \langle 1|U_1|0\rangle
    (x'(y')^*-g\alpha\beta^*)
    &=& 0 \label{cases}
\end{eqnarray}
If we chose
\begin{equation}
    \left(\begin{array}{cc} e^{i\phi} & 0 \\ 0 & e^{-i\phi} \end{array}\right)
\end{equation}
with $\phi=0$ and $\phi=\pi/2$ (which means $g=1$) then we get two equations
and with the resulting condition
\begin{equation}
    |x'| = |\alpha| \;\; \mbox{and} \;\; |y'| = |\beta|
\end{equation}
Now we chose two matrices from the set B to determine $g$. From
equation (\ref{cases}) we then find that
\begin{eqnarray}
    a^*(1-g)|\beta|^2 + b({x'}^* y' - \alpha^*\beta g) = 0 \label{eq1}\\
    a(1-g)|\alpha|^2 - b^*({x'} {y'}^* - \alpha\beta^* g) = 0 \label{eq2}
\end{eqnarray}
as coefficients in front of $e^{i\phi}$ and $e^{-i\phi}$ have to vanish. As
$a,b$ and $g$ are fixed we can now only vary $\alpha$ and $\beta$. We know that
$|\bar\chi\rangle$ and therefore $g$ do not depend on the choice of $\alpha$
and $\beta$. To determine $g$ let us now chose a special case, namely
$\alpha=0$:\\
In that case we see from (\ref{eq2}) that
\begin{equation}
    b^*{x'} {y'}^* = 0
\end{equation}
and therefore from (\ref{eq1}) we find
\begin{equation}
    a^*(1-g)=0
\end{equation}
Now we can consider three cases:\\
a) $a\neq 0,b\neq 0$: Then $g=1$.\\
b) $|a|=1$: Then $g=0$, but in that case the sets A and B are identical and
we already know the optimal protocol.\\
c) $|b|=1$: Again we know the optimal protocol already.\\

Therefore, we only need to consider the case where $|g|=1$ and $a\neq 0,b\neq
0$. Then we have that ${x'}^* y' = \alpha^*\beta$. Dividing both sides by
$|x'|^2$ gives
\begin{equation}
    \frac{y'}{x'} = \frac{\alpha^*\beta}{|x'|^2} =
    \frac{\alpha^*\beta}{|\alpha|^2}= \frac{\beta}{\alpha}
\end{equation}

This implies
\begin{equation}
    |0\rangle|\Phi_0\rangle + |1\rangle|\Phi_1\rangle = \frac{1}{\beta}
    \left(\alpha|0\rangle+\beta|1\rangle\right)|\Phi_1\rangle
\end{equation}
As the state $|\psi\rangle$ is general, this implies that $G_1$ is a state
transfer from Bob to Alice and the resource cost is 2 bits from Bob to Alice.
if we only wish to expend 1 bit from Bob to Alice, then this is not a valid
option and we can then therefore say that in general $G_1$ will produce an
entangled state.\\

ii) Now we can assume that there is a state $|\psi\rangle$ such that $G_1$
acting on $|\chi\rangle|\psi_1\rangle$ generates an entangled state. Let us
now make a basis change such that we can write
\begin{eqnarray}
    G_1|\chi\rangle|\psi_1\rangle &=& r|0\rangle|0\rangle +
    s|1\rangle|1\rangle \label{state}
\end{eqnarray}
Now we have to show that when a $U$ from any of the sets A or B acts on one half
of the state (\ref{state}), it is not possible to find a $G_2$ (unless either
$|a|=1$ or $|b|=1$) such that
\begin{eqnarray}
    G_2\left( U\otimes {\bf 1} (r|00\rangle + s e^{i\phi}|11\rangle \right) =
    U \left(\begin{array}{cc} 1 & 0 \\ 0 & e^{i\phi}
    \end{array}\right)|\psi\rangle|\chi_{U}\rangle  \label{line1} \\
    G_2\left( r|00\rangle + s e^{i\eta}|11\rangle \right) =
    U \left(\begin{array}{cc} 1 & 0 \\ 0 & e^{i\eta}
    \end{array}\right)|\psi\rangle|\chi\rangle \label{line2}
\end{eqnarray}
Firstly we note again that the state $|\chi\rangle$ cannot depend on $\phi$ or
$\eta$ as otherwise the trafo $G_2$ would not be linear. However, it may depend
on the choice of $U$. Now let us take the scalar product between Eq \ref{line1}
and Eq. \ref{line2}. Again $G_2$ drops out due to its unitarity and if we use that
$g=\langle\chi_{U}|\chi\rangle=1$ to find
\begin{equation}
    |r|^2 a + |s|^2 e^{i(\phi-\eta)} a^* = (a|\alpha|^2+a^*|\beta|^2e^{i(\phi-\eta)}
    + b\alpha^*\beta e^{i\phi} -
    b^*\alpha\beta^* e^{-i\eta})
\end{equation}
or
\begin{equation}
    (|r|^2-|\alpha|^2) a + (|s|^2-|\beta|^2) e^{i(\phi-\eta)} a^* -  b\alpha^*\beta e^{i\phi}
    +  b^*\alpha\beta^* e^{-i\eta} = 0
\end{equation}
for all $\phi,\eta$. This implies, that
\begin{equation}
    |r|^2-|\alpha|^2 = |s|^2-|\beta|^2 = 0\; \mbox{and} \; b\alpha^*\beta=0
\end{equation}
Because both $r$ and $s$ are non-zero, we find that also $\alpha$ and $\beta$
are non-zero, which implies that $b=0$ \cite{entangle}. Therefore the only two possible values
for $a$ and $b$ are $|a|=1$ and $|b|=1$ and this finishes the proof.
\section{Trade-off in resources}
The results of the previous section allows us to establish the uniqueness of
the two teleportable sets which arise in section IV as the two possible cases
were the transmission of just one bit from Alice to Bob was sufficient to design a
protocol for perfect remote implementation. It should be stressed that the
procedure works independently of to which particular subset the transformation
belongs to. Imagine now that Alice is given the promise that her apparatus can implement
transformations within a particular subset, for instance,
any
unitary transformation that commutes with the action of the Pauli operator
$\sigma_z$. In other words, she is provided with a machine that can implement
arbitrary rotations around the z-axis. As before, the aim is to implement remotely any
such transformation on Bob side, provided that he may hold a qubit state in an
arbitrary state $\ket{\psi}_B$. We
will show in the following that a variation of the protocol discussed in
section III allows to implement an arbitrary rotation on Bob side consuming just
1 e-bit and 1 c-bit in each direction. In contrast, BQST would consume 1 e-bit
and 2 c-bits per state teleportation step.
We start with Alice and Bob sharing an e-bit that for concreteness we assume to be the
maximally entangled state $\ket{\phi}_{AB}^+$. Bob holds a qubit system in an
arbitrary state $\ket{\phi}_B=\alpha \ket{0} + \beta \ket{1}$. We carry on the
same local operations on Bob side described in subsection III.A, that is, a
controlled-NOT between Bob's qubits with the unknown state acting as the
control qubit
followed by a projective measurement of the target qubit in the computational
basis. This sequence consumes 1 c-bit from Bob to Alice and ends up with both
parties sharing the, in general partially entangled, state $\alpha \ket{00} +
\beta \ket{11}$. Alice now applies the operation $U_c$
onto her qubit followed by a Hadamard transformation. No extra shared entanglement will be
required. The global (unnormalized) state of the distributed system after this action can be written as
\begin{eqnarray}
\ket{\lambda}_{AB} &=& \alpha \,a (\ket{0}_A + \ket{1}_A) \ket{0}_B + \beta \,a^\star
(\ket{0}_A - \ket{1}_A) \ket{1}_B \nonumber \\
&=& \ket{0}_A (\alpha \,a \ket{0}_B + \beta \,a^\star \ket{1}_B) +
\ket{1}_A (\alpha \,a \ket{0}_B - \beta \,a^\star \ket{1}_B)\nonumber \\
&=& \ket{0}_A (\alpha U(\ket{0}_B) + \beta U(\ket{1}_B))+
\ket{1}_A (\alpha U(\ket{0}_B) - \beta U(\ket{1}_B))
\end{eqnarray}
A projective measurement in the computational basis on Alice's side
yields a collapsed
state on Bob side which is either the correct transformed state by $U_c$, whenever the
measurement outcome is $\ket{0}_A$, or a state that can be locally transformed into the
correct one. If the measurement outcome is $\ket{1}_a$, all Bob
has to do is applying the correcting operation $\sigma_z$. Bob needs to know
the measurement outcome of Alice's measurement and therefore a further c-bit is
consumed in the second part of the protocol.
Identical results follow
if Alice is given the promise that the transformation $U$ anti-commutes with
$\sigma_z$. The only difference is that Bob gets the correct transformed state
via the application of  different correction steps, $\sigma_x$ for outcome $\ket{0}_A$
and $\sigma_z \sigma_x$ for outcome $\ket{1}_A$.\\
The explicit construction of a protocol that achieves the remote implementation
of any unitary operation of the form $U_c$ or $U_a$ proves that consuming 1
e-bit and 1 c-bit in each direction is sufficient. The necessity can be derived
from the following argument. Assume that we can teleport any transformation $U$
which either commutes or anticommutes with $\sigma_z$. We can then assume that
we could also implement any controlled-$U$ of that form, and in particular we
could implement a controlled-NOT operation. But it is known that the non-local
implementation of a controlled-NOT requires 1 e-bit and two classical bits in
each direction \cite{poli}, therefore the protocol we have described is
optimal.
\section{Conclusions and prospects}
We have analyzed the problem of performing quantum remote control on a qubit.
The principles of non-increase of entanglement under LQCC and the impossibility
of superluminal communication allow to establish lower bounds on the amount of
entanglement and the classical communication cost of a universal remote
control protocol: Alice and Bob need to share at least two e-bits and need to
communicate no less than two c-bits from Alice to Bob and one c-bit from Bob to Alice.
This asymmetry in the communication cost opens the possibility of a different
strategy than resorting to bidirectional state teleportation (BQST). While the
protocol cannot work perfectly for an arbitrary transformation on a qubit, we
have shown here that there are restricted sets of {\em teleportable}
operations, i.e., operations that can be implemented remotely consuming less
overall resources than BQST. Remarkably,
up to a local change of basis, only two {\em teleportable} subsets
exit: Arbitrary rotations around a fixed direction $\vec{n}$ or
rotations by a fixed angle around an arbitrary direction lying in
a plane orthogonal to $\vec{n}$.\\
We will finish by describing a possible experimental scenario
where the ideas we have developed could be demonstrated. From the practical
point view, the most challenging requirement arises from the distribution of a
highly entangled state between two remote parties. Nevertheless, theoretical
proposals have been made for establishing a maximally
entangled state of two trapped ions surrounded by an optical cavity \cite{innsbruck}. Let us
then assume that a maximally entangled state can be created using these
techniques. In addition, Bob's cavity holds a second ion initially prepared in
a state that for simplicity we will suppose to be an equally weighted superposition of levels
$\ket{0}$ and $\ket{1}$. Transformations which either commute or anticommute with the action
of $\sigma_z$ can be easily
realized by means of irradiating Alice's particle with laser light with a suitable value of
the ratio $\Delta/\Omega$, where $\Delta$ is the detuning from the atomic
transition $\ket{0}\longrightarrow \ket{1}$
and $\Omega$ the laser Rabi frequency. Applying the protocol described in
section VII leads to Bob holding a state of the form
\begin{equation}
\ket{\psi}_B= \frac{1}{2}(\ket{0}+e^{-i \vartheta}\, \ket{1}),
\end{equation}
where $\vartheta$ will be a function of the laser parameters. Therefore, a
subsequent measurement of Bob's particle in the $\ket{\pm}$ basis yields a
probability for the ion to be found in the $\ket{+}$-state
\begin{equation}
P_{\ket{+}}=\frac{1 + \cos \vartheta}{2}.
\end{equation}
In other words, under repeated measurements following laser irradiations of
different duration on Alice's side, Bob's particle, in a remote location, will exhibit Ramsey fringes.
This effect is a nice illustration of how quantum non-locality can be exploited and should
lie among the near future experimental capabilities in quantum optics. Applications in
quantum communication protocols are foreseeable.\\

{\em Acknowledgements:} We acknowledge discussions with R. Ratonandez and B.
Reznik. This work was supported by the UK Engineering and Physical Sciences
Research Council (EPSRC), by the EQUIP project of the European Union and the
European Science Foundation Programme on 'Quantum Information Theory and
Quantum Computing' and by the ESF-QIT
conference 'Quantum Information Theory: Theory, Experiment and Perspectives'
in Gdansk, 2001.\\

Note: After this work was completed, we have learned of related results
obtained independently by B. Reznik (unpublished) and Chui-Ping Yang
and J. Gea-Banacloche, quant-ph/0107100.

\section{appendix}

Here we give the proof that states $\ket{\psi}$ and
$\ket{\psi_\perp}$ can be found to satisfy Eqs.\ (\ref{Upsi}) and
(\ref{Upsi_perp}) for $\ip{\psi|\psi_\perp}=0$ and
$\ip{\phi|\phi^\prime}\ne 0$. We drop the superscripts $(n), (m)$
from the operators in these equations and diagonalize the
unimodular product $U^\dagger_2 U_1$:
\[
  U^\dagger_2 U_1 \ket{\lambda_\pm} = e^{\pm i\lambda}\ket{\lambda_\pm}\ .
\]
Note that $\lambda\ne 0, \pi, 2\pi...$ for otherwise $U^\dagger_2
U_1=\pm(\ket{\lambda_+}\bra{\lambda_+}+\ket{\lambda_-}\bra{\lambda_-})$
which is proportional to the identity, and so the operators would
be trivially related $U_2=\pm U_1$ forcing $\ket{\phi}$ and
$\ket{\phi^\prime}$ to be orthogonal. Let
\begin{eqnarray}
  \ket{\psi} &=& (\ket{\lambda_+}+\ket{\lambda_-})/\sqrt{2}
    \nonumber\\
  \ket{\psi_\perp} &=& (\ket{\lambda_+}-\ket{\lambda_-})/\sqrt{2}
  \label{exp_psi_perp}
\end{eqnarray}
then
\begin{eqnarray*}
 U^\dagger_2 \ket{\phi}&=& U^\dagger_2 U_1 \ket{\psi}\\
                       &=&(e^{i\lambda}\ket{\lambda_+}
                           +e^{-i\lambda}\ket{\lambda_-})/\sqrt{2}
\end{eqnarray*}
from which we find
\begin{equation}
  \ket{\phi}=U_2 (e^{i\lambda}\ket{\lambda_+}
                           +e^{-i\lambda}\ket{\lambda_-})/\sqrt{2}
  \label{U2_1}
\end{equation}
As before we also have
\begin{equation}
  U_2\ket{\psi_\perp}=\ket{\phi^\prime}\ .
  \label{U2_2}
\end{equation}
Thus from Eqs.\ (\ref{U2_1}), (\ref{U2_2}) and
(\ref{exp_psi_perp}) we get
\begin{eqnarray*}
  \ip{\phi^\prime|\phi}
    &=& \bra{\psi_\perp}U^\dagger_2 U_2(e^{i\lambda}\ket{\lambda_+}
                           +e^{-i\lambda}\ket{\lambda_-})/\sqrt{2}\\
    &=&i \sin(\lambda)
\end{eqnarray*}
which is nonzero (because $\lambda\ne 0, \pi, 2\pi...$).\\



\begin{thebibliography}{99}
\bibitem{basis} The final correction operation $V$ can be written as $V = W \,
\sigma_z \, W^{-1}$ where $W$ is a generic uni-modular transformation.
Eq.(\ref{corr}) can now be re-written as $\sigma_z \, W^{-1} U = e^{i \gamma}
\, W^{-1} U \, \sigma_z$ and the two possible solutions are given by operations
of the form $W U_c$ or $W U_a$. However, this amounts to the trivial degree of
freedom of performing a basis change.
\bibitem{entangle} Note that it was used explicitly that $G_1$ generates an entangled
state, ie that $r$ and $s$ are non-zero.
\end{thebibliography}
\end{document}